# Offloading MPI Parallel Prefix Scan (MPI_Scan) with the NetFPGA

Omer Arap   Martin Swany
Center for Research in Extreme Scale Technology
Indiana University
Bloomington, IN 47405
{omerarap,swany}@crest.iu.edu

*Abstract*—Parallel programs written using the standard Message Passing Interface (MPI) frequently depend upon the ability to efficiently execute collective operations. `MPI_Scan` is a collective operation defined in MPI that implements parallel prefix scan which is very useful primitive operation in several parallel applications. This operation can be very time consuming. In this paper, we explore the use of hardware programmable network interface cards utilizing standard media access protocols for offloading the `MPI_Scan` operation to the underlying network. Our work is based upon the NetFPGA – a programmable network interface with an on-board Virtex FPGA and four Ethernet interfaces. We have implemented a network-level `MPI_Scan` operation using the NetFPGA for use in MPI environments. This paper compares the performance of this implementation with MPI over Ethernet for a small configuration.

## I. INTRODUCTION

The *Parallel prefix scan* or *sum* is a collective operation which is remarkably very versatile primitive in various parallel algorithms as presented in [8]. The *Message Passing Interface* (MPI) also defines scan in two different collective operations: `MPI_Scan` as an inclusive scan and `MPI_Exscan` as an exclusive scan operation. Over the past years, various algorithms have been proposed to enhance the performance of parallel prefix scan operation. In general, offloading collective logic is a common technique in various platforms to improve MPI performance. In this paper, we study the effect of offloading on performance of the parallel prefix scan operation for various algorithms in MPI programming model.

Offloading collective operation to the network substrate is a common technique in various platforms to improve MPI performance. The host CPU offloads the collective to the network interface card (NIC) and waits for the result of the collective operation. If the operation is implemented in a non-blocking fashion, the CPU can continue execution and waits for the collective result at some other point throughout the execution. Otherwise, it stalls the execution and waits for the result to be returned to the application. In the past, offloading different collective operations defined in MPI have been an attractive research area, and significant performance gains have been shown. In fact, we also studied `MPI_Barrier` implementation on a NetFPGA cluster and presented results in our previous work [6]. To the best of our knowledge, this is the first attempt to offload `MPI_Scan`.

We use the NetFPGA platform, which includes a Field Programmable Gate Array (FPGA) component, allowing the "programming" of custom digital logic in hardware [14]. This infrastructure can provide high performance, efficient realization of performance critical functionality. The NetFPGA embeds this in a NIC, providing an ideal infrastructure for message passing applications.

In this paper, we propose a parallel prefix scan framework for MPI utilizing the standard infrastructure from the NetFPGA platform and using standard protocols such as UDP, IP and Ethernet. The unique contributions of our work are as follows:
- The design relies on the standard NetFPGA driver and there is no need to change anything in the OS. We incorporate some simple changes in the user-level code, utilizing the Open MPI [3] library to generate the packets that the NetFPGA recognizes and processes.
- All of our additional hardware modules live in the user-data-path [5], as recommended by the NetFPGA user community. Therefore, it is self-describing and could be extended by someone who is familiar with the NetFPGA environment.
- We provide different algorithm selection at the hardware level. Therefore MPI runtime can make an intelligent selection of algorithms based on the underlying network topology.
- Our work does not require a separate control network for parallel prefix scan operation as it can perform scan operation and produce the outcome for each rank on the network where the data also flows.

The remainder of this paper is organized as follows: Section II provides background information. Section III outlines the implementation details with the NetFPGA. Section IV presents performance evaluation of our design. Section V provides related work. Section VI offers discussion about our work and how it could be extended in the future, and finally Section VII concludes the paper.

## II. BACKGROUND

This section provides background for our work, describing the NetFPGA platform and parallel prefix scan algorithms used in major MPI suites [3] [2]. Additional related work is discussed in Section V.

## A. Parallel Prefix Scan

In this section, we explain the definition of parallel prefix scan. Let $p$ be the number of ranks in a communicator numbered from 0 to $p-1$, and let a sequence of $p$ elements $x_i$ with an associative, binary operation $\oplus$ be given. The inclusive parallel prefix operation computes for each rank $j$, $0 \leq j < p$ the $\bigoplus_{i=0}^{j} x_i = x_0 \oplus x_1 \oplus ... \oplus x_j$. The exclusive parallel prefix operation is somewhat similar to the inclusive version, except that the rank $j$ does not add its local value to its partial sum.

## B. Scan Algorithms

Even though there are variety of parallel prefix scan algorithms, we selected the algorithms that are currently employed in Open MPI and MPICH. In this section, we only present communication pattern of the algorithms and will leave the implementation details in the next section. For simplicity, we define $p$ as the number of ranks and $p$ is in powers of two.

*1) Sequential Algorithm:* This algorithm is currently used in Open MPI. It is hard to claim that it is a parallel algorithm since starting from rank 0, each rank sends its partial result to the next rank, after receiving the partial result from the previous rank. The algorithm runs in $p$ steps.

*2) Recursive Doubling:* This algorithm is also called *Naive Algorithm* and it is the selected algorithm by MPICH MPI suite. The recursive doubling algorithm requires $\log_2 p$ steps to reach the end of the algorithm. In step $k$, where $k = 0, ..., \log_2 p - 1$, a rank $j$ exchanges the cumulative sum so far with the rank $j$ ^ $2^k$ ( ^ is "bitwise XOR"). Rank $j$ also needs to track the incoming messages and decide if they need to be added to its partial sum. To decide that, it checks if $j \& k \neq 0$ (& is "bitwise and"). Then, it sums the received message with its partial sum and updates its partial sum. Unlike the sequential algorithm, this algorithm also involves implicit process synchronization. After each message exchange, the algorithm requires ranks to buffer the messages in order to calculate final outcome after the message exchanges are over.

*3) Binomial Tree:* This algorithm is first presented by [9]. It is called *binomial tree algorithm*, because the ranks' communication pattern maps to a binomial tree. The binomial tree have 2 separate phases: *up-phase* and *down-phase*. Both phases complete in $\log_2 p$ steps. For the up-phase, in step $k$, where $k = 0, ..., \log_2 p$; rank $j$ that satisfies the condition $j \& (2^{k+1} - 1) = 2^{k+1} - 1$ receives a partial sum from rank $j - 2^k$ where $0 \leq j - 2^k$. Then, it begins to wait for the down-phase. In the up-phase, a leaf rank only sends its data to its parent and keeps waiting for the down-phase. In each step, an internal rank either receives a partial sum from one of its children or generates its partial sum and sends it to its parent after receiving all the partial sums from its children. The receiving ranks, which are internal and root ranks, add the partials results received from their children, and after step $k$ they have the partial result of ranks $[j - 2^{k+1} + 1..j]$. When the root receives messages from all of its children, the down-phase begins. In the down-phase, the steps go from $lg_2 p$ to 1. Rank $j$ that satisfies the condition $j \& (2^k - 1) = 2^k - 1$ sends its partial sum to rank $j + 2^{k-1}$ so that the receiving rank can produce its local final result.

## C. NetFPGA

The NetFPGA is an open-source programmable network device and software ecosystem developed at Stanford University with support from the NSF and industry partners. The NetFPGA was designed to facilitate high performance networking research and instruction in network hardware design. It has been widely used by researchers and educators since its release in 2005. In 2009, an effort began to create a next generation NetFPGA that could satisfy increasing performance needs, providing 10Gb/s Ethernet. Increasing the performance dramatically in an open, extensible platform is challenging and the software for this system is currently in beta release status.

The first generation NetFPGA is a PCI card featuring a Xilinx Virtex-II Pro FPGA running at 125 MHz, 4 1Gb/s Ethernet ports, 4.5 MB of static RAM and 64MB of DDR2 dynamic RAM. While the NetFPGA-10G effort follows much of the original NetFPGA design philosophy, the hardware and architecture are different. The NetFPGA-10G is a PCI-Express board with a Xilinx Virtex-5 FPGA with a 160Mhz clock, four 10 Gb/s Ethernet ports, 27 MB of SRAM and 288 MB DRAM. Our work is based upon the original NetFPGA, which uses 1Gb/s Ethernet.

## III. IMPLEMENTATION

Our FPGA node design is derived from the *reference NIC* implementation distributed with the NetFPGA package. The host communicates with our offload engine through a UDP socket – operating system support for such sockets is part of the standard package. The NF_Scan implementation consists of sending a specially crafted UDP message, and then blocking until a final outcome of the scan is received. An added feature of building our implementation upon the NetFPGA reference NIC is that our node maintains the ability to forward standard IP packets.

The simplicity of the host interface belies the complex task that the scan node must perform. The NetFPGA tracks outstanding requests by storing the various MAC, IP addresses, checksum and UDP header fields. These are later used to generate a result message to inform the host. The result packet must arrive user-space travelling up to the protocol stack. Therefore, it must be properly formed, so that none of the layers prevent packet to be processed by the application layer.

## A. Packet Format

Our design is intended to support a variety of collective operations and algorithms. We use the packet format presented in Figure 1 to inform the NetFPGA about which network level state machine to utilize in order to execute the collective operation. Not all the fields are necessary for every type of collective operation. For the parallel prefix scan, depending on the algorithm, almost all fields could be necessary. *comm_id* is a unique identifier of the communicator to distinguish the states of simultaneous collective operations that might be running on the network. This feature has not been implemented yet, and left as the future work. *comm_size* denotes the number of participating processes to the collective operation. *coll_type* is the collective operation type and it is enumeration of MPI_Scan for this work. *algo_type* is the algorithm for the collective operation that the NetFPGA will run. *node_type* is the rank's specific role in that algorithm. For example, in a binomial tree algorithm the highest rank becomes the root of the tree and it employs a different state machine than internal and leaf nodes. The *node_type* could be derived from the *rank* and *comm_size* fields in the hardware, but for simplicity, we let the software assign node roles in advance, and let the NetFPGA run the algorithm based on the assigned *node_type* role. *msg_type* field is needed when NetFPGAs communicate between each others, and notify what the packet means or could be thought as the metadata. *root* is not used for MPI_Scan and it denotes the target rank for collectives like MPI_Reduce. *operation* enumerates the reduction operation that should be performed on the data, *data_type* denotes the type of the data elements and finally *count* is the number of elements.

| | | | | | | | | | | | | | | | |
|---|---|---|---|---|---|---|---|---|---|---|---|---|---|---|---|
| 0..3 | 4..7 | 8..11 | 12..15 | 16..19 | 20..23 | 24..27 | 28..31 | 32..35 | 36..39 | 40..43 | 44..47 | 48..51 | 52..55 | 56..59 | 60..63 |
| dst_MAC | | | | | | | | | | | | src_MAC_1 | | | |
| src_MAC_2 | | | | type | | | | ver | IHL | Diff_Serv | | | | | |
| Total_Length | | | | Identification | | | | flags | | fragment_offset | | TTL | | Protocol | |
| Hdr_Cksum | | | | src_IP | | | | | | | | dst_IP_1 | | | |
| dst_IP_2 | | | | UDP_Source_Port | | | | UDP_Dest_POrt | | | | Length | | | |
| UDP_checksum | | | | comm_id | | | | comm_size | | | | coll_type | | | |
| algo_type | | | | node_type | | | | msg_type | | | | rank | | | |
| root | | | | operation | | | | data_type | | | | count | | | |

Fig. 1: Fields and structure of an actual NetFPGA collective operation offload packet

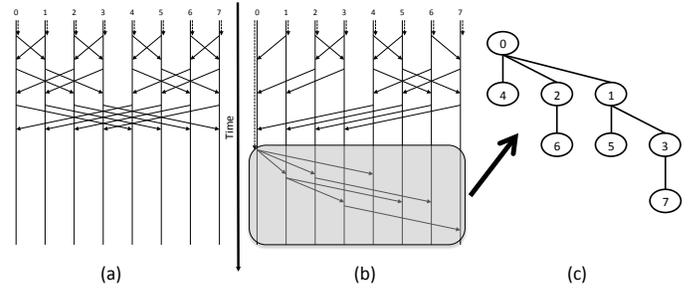

Fig. 2: Recursive doubling pattern: a. Standard recursive doubling pattern b. One late rank c. Corresponding binomial tree

### B. Sequential Algorithm

Even though algorithmically sequential algorithm is simple, it still needs a special handling when it is offloaded to the NetFPGA. When back-to-back MPI_Scan calls occur, the NetFPGA must handle this very carefully because the NetFPGA have limited resources. In the sequential algorithm, all ranks return in different times. For example, in a scenario of 4 processes, rank 0 might call MPI_Scan earlier than other ranks. Since rank 0 does not wait for receiving any data from other processes, it can return and throughout the execution, it can simply call MPI_Scan again. However, rank 1 might have not even called the first MPI_Scan, and it requires special logic to distinguish MPI_Scan requests. However, no matter how much we try to buffer outstanding MPI_Scan requests, the resources are limited. Therefore, we introduce an acknowledgment mechanism for this algorithm, so that rank $j$ does not immediately return after it generates its final outcome. It waits an acknowledgment packet from rank $j + 1$. The NetFPGA of rank $j+1$ sends an acknowledgment packet to the NetFPGA of rank $j$ after it receives MPI_Scan request from its host and the packet from rank $j$. With this technique, even though rank $j + 1$ waits an acknowledgment from rank $j + 2$, it can simply require a single buffer to buffer an outstanding packet received from rank $j$.

### C. Recursive Doubling Algorithm

The recursive doubling algorithm is a popular algorithm in other collective operations when the message size is small. We have implemented this algorithm in our previous works for MPI_Barrier [6] and MPI_Allreduce [7]. The ability to reconfigure the network provided more optimizations to the recursive doubling algorithm by using the NetFPGA's feature to multicast packets to selected peer NetFPGAs. However, in those types of collective operations, the outcome of the collective is the same for all the participating ranks. It is not the case for MPI_Scan because every rank receives a partial sum and it is unique to each rank. Therefore, this does not let us utilize fully optimized version of recursive doubling algorithm in offloaded version of MPI_Scan. The main idea in our previous work [7] is that in case there is a late rank arriving to the collective operation, it can generate a cumulative result. The butterfly topology running the recursive doubling algorithm adopts itself to a binomial tree with late rank being the root of the tree. That situation is depicted in Figure 2.

Even though the entire idea does not work for MPI_Scan, there is still the possibility to take advantage of multicasting between two consecutive recursive doubling stages if a rank $j$ arrives late and its peer in stage $k$ has already sent its message to rank $k$. Each rank is required to buffer incoming data from its peers if it uses received data in the final outcome of the operation. In the scenario depicted in Figure 3.a rank 0 and 1 arrives to the collective point at the same time interval, and the algorithm runs normally. In Figure 3.b, rank 1 arrives later than it received a message from rank 0. It generates a cumulative partial data of rank 0 and 1 and sends it to rank 0 and 3 via multicasting. Therefore, it does not need to generate separate messages for both ranks. It also tags the message which indicates that the packet contains partial result of rank 0 and 1. When rank 0 receives this packet, since it has already cached its own partial result, it can subtract that result from the cumulative message and derives rank 1's message locally. This optimization does not work for all data types and operations. However it is perfect for data type MPI_INT performing MPI_SUM, since subtraction is inverse of addition

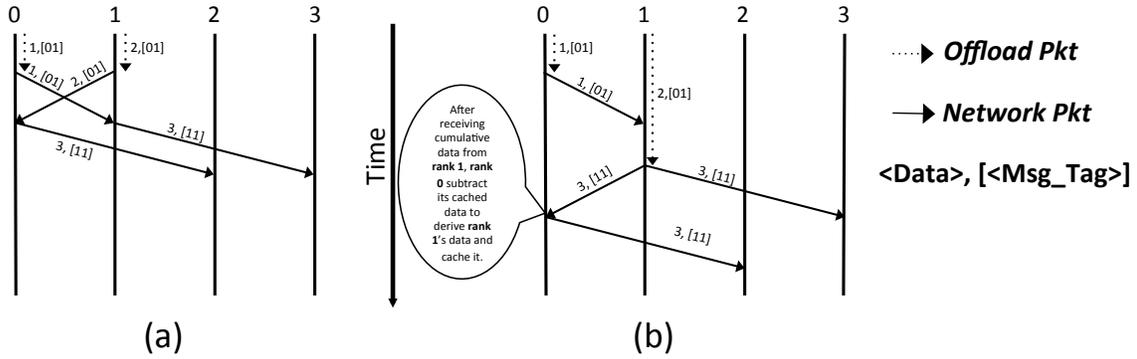

Fig. 3: Optimization to the recursive doubling algorithm with message tagging and multicasting.

and we do not need extra cycles to perform subtraction while streaming the data.

### D. Binomial Tree Algorithm

In this algorithm, ranks are organized as a binomial tree. The binomial tree is also popular algorithm for various other MPI collective operations such as `MPI_Allreduce`, `MPI_Barrier` etc. However, binomial tree algorithm in `MPI_Scan` totally differs from other collectives since a source rank sends different data to different ranks unlike other collective operations. For example, in `MPI_Allreduce` the accumulated data is gathered in the root rank and then multicasted to its children. Heterogeneity of the final outcome in `MPI_Scan` breaks this rule, and it is not possible to take advantage of the NetFPGA network's multicasting capability in this algorithm. The communication pattern is described in previous section after the offload request has been issued to the NetFPGA. The NetFPGA have preallocated buffers to cache children's messages during the up-phase of the algorithm. During the down-phase it fetches the data from those caches and generates partial sums for its children and sends them back-to-back. Since messages are generated at the hardware, it happens at the line rate and does not need to fetch anything from the host's memory.

## IV. EVALUATION

### A. Experimental setup and results

Our experimental setup consists of 8 NetFPGAs in hosts with Intel(R) Core i5-2400 at 3.10GHz CPUs, 4GB RAM, and a dual Gigabit Ethernet NIC. The NetFPGA ports were directly connected to the each other establishing a tested topology. In this paper, we present micro-benchmark results obtained running modified version of OSU Micro-Benchmark Suite [4] for `MPI_Scan`. In addition, we are going to describe how we can precisely time the NetFPGA operations after we offload the collective to the NetFPGA network.

The benchmark is configured to run 10 million back-to-back `MPI_Scan` calls and average and minimum latency results are recorded. Figure 4 shows the average latency of a single `MPI_Scan` operation for different message sizes observed in 8-node settings for various algorithms. Figure 5 also presents minimum latency observed in those calls. The offloaded versions are denoted by the prefix *"NF_"* in these figures. We did not provide the results for binomial tree for software based implementation since it produced the worst performance.

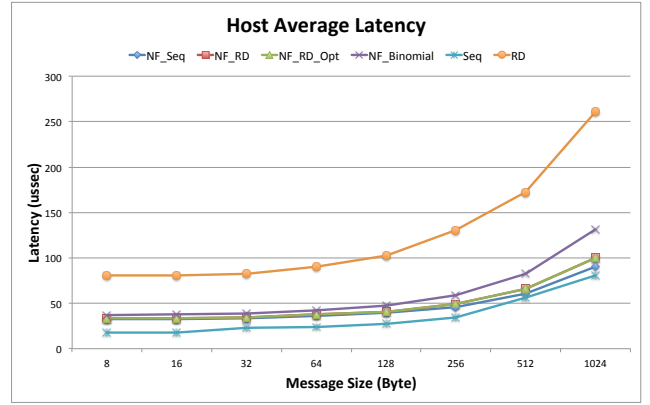

Fig. 4: Comparison of software based and offloaded versions of `MPI_Scan` in terms of average latency observed.

According to these graphs, software version of the sequential algorithm produces the minimal latency. Essentially, the sequential algorithm does not involve implicit synchronization among participating processes. Therefore, once a process produces its partial sum, it simply returns and continues to its execution. For example, if rank $j$ has already buffered the partial sum from rank $j-1$ before it calls `MPI_Scan`; it suffers almost zero latency for calling `MPI_Scan` since it immediately adds its local data to the received partial sum, forwards the result to the rank $j+1$ and simply returns. The data transfer is handled in another layer of the MPI stack. Therefore, the host process does not must wait for the transfer to be completed. With this keeping in mind, some ranks suffer very less latency in the software version of sequential algorithm and this results in low average latency.

On the other hand, in the offloaded version, host process needs to interact with the NetFPGA 2 times: one for offloading the request and the other for receiving the final result. In addition, because of the limited NetFPGA resources and avoiding from back-to-back conflict, the NetFPGA employs explicit acknowledgement mechanism which introduces extra latency. The acknowledgements are present in the software version also, but they are handled by the TCP and host process has direct involvement in acknowledgement mechanism because different MPI module is responsible for that. For the algorithms that enforce synchronization, offloaded versions beat the software based implementations by significant values. Even though sequential algorithm produces the best results in 8-node settings, it is not scalable algorithmically and would produce significant performance degradation on big clusters.

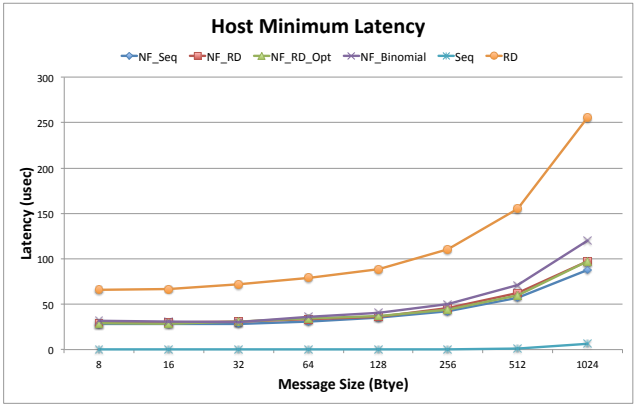

Fig. 5: Comparison of software based and offloaded versions of `MPI_Scan` in terms of minimum latency observed.

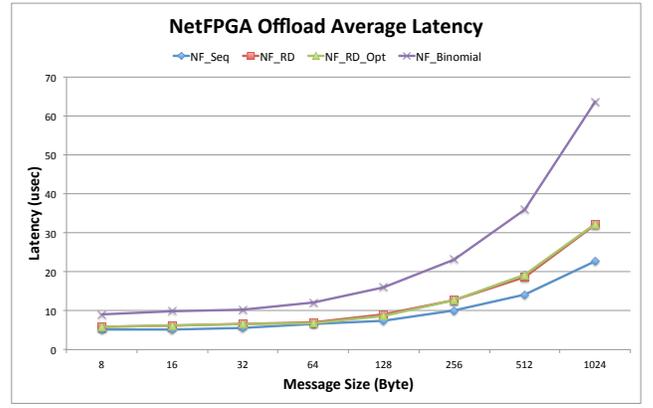

Fig. 6: Comparison of different `MPI_Scan` algorithms for offloaded version in terms of average latency observed after the offload is issued to the NetFPGA.

As we mentioned in the introduction, the NetFPGA's host driver is not optimized unlike modern interconnects. It does not employ techniques such as zero-copy, interrupt coalescing, pre-allocated packet buffers, and memory registration to reduce the latency of offload process between the host and the NetFPGA. The NetFPGA has been considered by the NetFPGA community as a networking device that could be utilized at the core of the network as a special purpose switch or a router. Therefore, its DMA engine and host driver did not attract attention to get optimized so far. We also did not focus on optimizing these components. Therefore, we were interested in measuring the latency of the `MPI_Scan` after it is offloaded to the NetFPGA network.

The NetFPGA has 125MHz clock which enables us to create an 8*ns* resolution timer. We initialize a 64-bit counter once the design (code) is loaded on the NetFPGA and the counter is incremented in every rising edge of the clock. We also create two 64-bit timestamp registers to track the offload and release time of the collective operations. *Offload time* is recorded once the initial packet for the collective is received from the local host process. When the collective algorithm reaches the completion or release state, and sends the final outcome of the collective to the local host process, the *release time* is recorded. The difference between the *release time* and *offload time* is the time elapsed in collective communication for that rank in the network. The *elapsed time* is attached to the collective result packet for our modified benchmark for further analysis. The average latency introduced by the NetFPGA network per host after the operation is offloaded to the NetFPGA network is presented in Figure 6. We also track the minimum observed latency for these measurements and those results are presented in Figure 7.

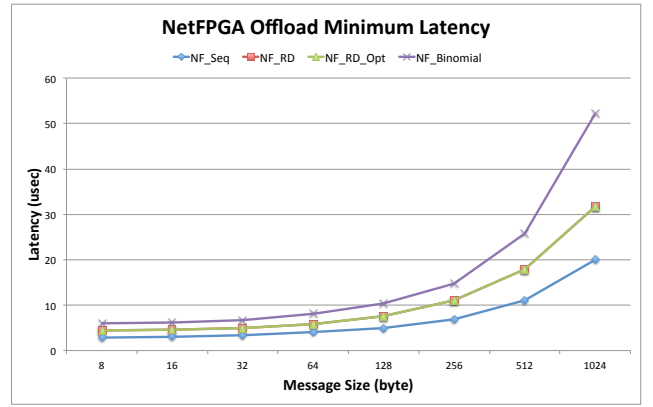

Fig. 7: Comparison of different `MPI_Scan` algorithms for offloaded version in terms of minimum latency observed after the offload is issued to the NetFPGA

## V. RELATED WORK

Our work in this paper is related to the very broad area of high performance computing. Parallel prefix scan operations have been studied for special purpose hardware and especially in FPGAs in the past. Park and Dai [15] demonstrate the design of a reconfigurable parallel prefix scan hardware on FPGAs by employing pipelined dataflow algorithm. Unlike our work, this study does not focus on MPI related applications and provides a platform where the entire scan operation is run on a single FPGA. Similarly, Vitoroulis and Al-Khalili [17] also study different hardware level implementation approaches of parallel prefix adders on a single FPGA. [13] is another example of parallel prefix implementation at the hardware level employing a single FPGA focusing on the area requirements and the critical path delay of parallel prefix hardware designs. Harris et al. [12] summarize parallel prefix scan algorithms for GPU based environments and provide implementation details in CUDA.

Sanders and Träff [16] provide a great survey of parallel prefix scan algorithms that are suitable for MPI based environments. We benefit from their work by selecting the algorithms presented in this work significantly. Our work differs from this effort significantly because our major focus is offloading the parallel prefix scan operation to the network substrate.

Collective offloading and hardware level support is highly related to our work and there have been a tremendous number of hardware level optimizations proposed in the past. CORE-Direct [1] technology by Mellanox is significantly related to our work which provides offloading of collective operations for InfiniBand based clusters. However, they do not provide any support for offloading `MPI_Scan`. The CORE-Direct feature was first presented by Graham et al. [10] demonstrating how task lists could be generated to implement offloaded versions of collective operations. In this work [11], they presented an offloaded recursive doubling task list for barrier synchronization.

## VI. Discussion and Future Work

Our design has obvious limitations, including manual configuration. We leave these to be addressed in future work. Moreover, even though we integrated our design into the Open MPI via simply replacing the included MPI_Scan, a more significant integration effort is necessary to preserve the architecture and semantics of Open MPI.

In our packet format we defined a field called *comm_ID*. However, it is not used in this design; the goal is to distinguish active collective operations, which may run on simultaneously for different MPI communicators. Each of the simultaneous collective operation will require a separate state machine. Therefore, in order to distinguish the states of active collective operations, we are planning to investigate the best way to store the *comm_ID* with their associated collective states. We are currently investigating approaches to store the (*comm_ID*, *collective_state*) tuples since the read and write operations for those tuples are going to be almost equal.

Moreover, we are planning to put hardware logic into the NetFPGA to learn the topology of the NetFPGA collective network and configure node roles as appropriate. This information will be propagated to the MPI environment, eliminating the hardcoding that comes with the current design and making it portable to other NetFPGA network configurations. We are also planning to achieve the self-configurability without changing any system level driver, and implementing the logic at the hardware and user-level.

## VII. Conclusion

In this paper, we have presented preliminary results using NetFPGAs to implement offloaded version of `MPI_Scan`. While the hardware designs presented have some limitations, the results provide strong evidence that this is likely to be a fruitful research domain. Limitations in our initial design include lack of mechanisms for failure recovery and the need for a pre-assigned node roles. Our plans include better and more robust implementations of `MPI_Scan` as well as other collective operatio mechanisms, performance evaluation on real parallel code, and integration with MPI libraries.

## References

[1] Core-direct the most advanced technology for mpi/shmem collectives offloads. http://www.mellanox.com/related-docs/whitepapers/TB_CORE-Direct.pdf.

[2] Mpich : High-performance portable mpi. http://www.mpich.org.

[3] Open mpi: Open source high performance computing. http://www.open-mpi.org.

[4] Osu micro-benchmarks 4.0. http://mvapich.cse.ohio-state.edu/benchmarks/.

[5] Reference router walkthrough. http://wiki.netfpga.org/foswiki/bin/view/NetFPGA/OneGig/ReferenceRouterWalkthrough.

[6] O. Arap, G. Brown, B. Himebaugh, and M. Swany. Implementing MPI Barrier with the NetFPGA. In *The 20th International Conference on Parallel and Distributed Processing Techniques and Applications*, Las Vegas, USA, July 2014. (in appear to).

[7] O. Arap, G. Brown, B. Himebaugh, and M. Swany. Software Defined Multicasting for MPI Collective Operation Offloading with the NetFPGA. In *Euro-Par 2014 Parallel Processing*. Springer, August 2014. (in appear to).

[8] G. E. Blelloch. Scans as primitive parallel operations. *Computers, IEEE Transactions on*, 38(11):1526–1538, 1989.

[9] G. E. Blelloch. Prefix sums and their applications. 1990.

[10] R. Graham, S. Poole, P. Shamis, G. Bloch, G. Bloch, H. Chapman, M. Kagan, A. Shahar, I. Rabinovitz, and G. Shainer. ConnectX-2 InfiniBand management queues: First investigation of the new support for network offloaded collective operations. In *Cluster, Cloud and Grid Computing (CCGrid), 2010 10th IEEE/ACM International Conference on*, pages 53–62, 2010.

[11] R. L. Graham, S. Poole, P. Shamis, G. Bloch, N. Bloch, H. Chapman, M. Kagan, A. Shahar, I. Rabinovitz, and G. Shainer. Overlapping computation and communication: Barrier algorithms and ConnectX-2 CORE-Direct capabilities. In *Parallel & Distributed Processing, Workshops and Phd Forum (IPDPSW), 2010 IEEE International Symposium on*, pages 1–8. IEEE, 2010.

[12] M. Harris, S. Sengupta, and J. D. Owens. Parallel prefix sum (scan) with cuda. *GPU gems*, 3(39):851–876, 2007.

[13] F. Liu, F. F. Forouzandeh, O. A. Mohamed, G. Chen, X. Song, and Q. Tan. A comparative study of parallel prefix adders in fpga implementation of eac. In *Digital System Design, Architectures, Methods and Tools, 2009. DSD'09. 12th Euromicro Conference on*, pages 281–286. IEEE, 2009.

[14] J. Lockwood, N. McKeown, G. Watson, G. Gibb, P. Hartke, J. Naous, R. Raghuraman, and J. Luo. Netfpga–an open platform for gigabit-rate network switching and routing. In *Microelectronic Systems Education, 2007. MSE'07. IEEE International Conference on*, pages 160–161. IEEE, 2007.

[15] J. H. Park and H. Dai. Reconfigurable hardware solution to parallel prefix computation. *The Journal of Supercomputing*, 43(1):43–58, 2008.

[16] P. Sanders and J. L. Träff. Parallel prefix (scan) algorithms for mpi. In *Recent Advances in Parallel Virtual Machine and Message Passing Interface*, pages 49–57. Springer, 2006.

[17] K. Vitoroulis and A. J. Al-Khalili. Performance of parallel prefix adders implemented with fpga technology. In *Circuits and Systems, 2007. NEWCAS 2007. IEEE Northeast Workshop on*, pages 498–501. IEEE, 2007.